# Study on Leveraging Wind Farm Reactive Power Potential for Uncertain Power System Reactive Power Optimization

Yu Zhou, *Student Member, IEEE*, Zhengshuo Li, *Member, IEEE*

*Abstract*—This paper suggests leveraging reactive power potential (RPP) embedded in wind farms to improve power system operational safety and optimality. First, three typical RPP provision approaches are analyzed and a two-stage robust linear optimization based RPP evaluation method is proposed. This approach yields an RPP range that ensures the security of wind farm operations under any realization of uncertainty regarding the wind farm. Simplified DistFlow equations are employed here for a compromise between computational accuracy and cost. Next, an uncertain RPP-involved reactive power optimization problem is introduced, through which system operators ensure system-wide security and optimality regarding the base case and against any possible deviation caused by uncertain lumped loads and renewable generation. Steady-state models of automatic generation control and local voltage control are also captured in this uncertain reactive power optimization, which is then transformed through Soyster's method into a deterministic optimization problem that is readily solvable. Case studies have conceptually validated that even with notable uncertainty, wind farms are still a competent reactive power resource providing considerable RPP. Also, simulation confirms positive and notable improvement of leveraging wind-farm RPP on system-wide operational security and optimality, especially for power systems with high wind penetration.

## Nomenclature

### A. Notations for Wind Farm RPP Evaluation Problem

| | |
|---|---|
| $p_{wj}, q_{wj}, v_{wj}$ | Active, reactive power, and squared voltage at node $wj$. |
| $\underline{q}_{w1}, \overline{q}_{w1}$ | Bounds for the RPP at the POI, i.e., node $w1$. |
| $\underline{Q}_w, \overline{Q}_w$ | Default limits for the reactive power at the POI. |
| $p_{wij}, q_{wij}, l_{wij}$ | Active, reactive power, squared current through line $wij$. |
| $p_{wij,0}, q_{wij,0}$ | Real-time measurements of $p_{wij}, q_{wij}$. |
| $r_{wij}, x_{wij}, t_{wij}$ | Resistance, reactance, and tap ratio regarding line $wij$. |
| $p_{wGj}, q_{wGj}, S_{wGj}$ | Active, reactive, and apparent power for DFIG $wGj$. |
| $p_{wGj,0}, \Delta_{wGj}$ | Forecasted power and possible deviation for DFIG $wGj$. |
| $q_{wSj}, q_{wCj}, S_{wSj}$ | Reactive power from SVG $wSj$ and capacitor banks $wCj$, and apparent power regarding SVG $wSj$. |
| $TR_{wij,\kappa}, CP_{wj,\kappa}$ | Possible tap ratios and capacitance of the capacitor banks. |
| $z_{wCj,\kappa}, z_{wTij,\kappa}, \iota$ | Auxiliary continuous variables regarding linearization. |
| $b_{wCj,\kappa}, b_{wTij,\kappa}$ | Auxiliary binary variables regarding linearization. |
| $V_{POI}^{set}, \beta$ | POI voltage setpoint and the preference weight. |
| $\mathcal{H}(j), \mathcal{T}(j)$ | The sets of the parent and child nodes of node $wj$. |
| $\mathbf{A}, \mathbf{B}, \mathbf{C}, \mathbf{D}, \mathbf{E}, \mathbf{F}$ $\mathbf{J}, \mathbf{R}, \boldsymbol{g}, \boldsymbol{h}, \boldsymbol{m}, \boldsymbol{r}, \boldsymbol{s}$ | The coefficient matrices (bold capitals) and vectors (in bold italics) in the canonical RO formulation. |
| $\omega, \vartheta, \phi, \eta, \boldsymbol{b}_1, \boldsymbol{b}_2$ | Auxiliary variables used in the solution method. |

### B. Notations for System-wide Optimization Problem

| | |
|---|---|
| $V_i, \theta_i, p_i, q_i$ | Voltage, phase angle, active and reactive power at bus $i$. |
| $\boldsymbol{V}, \boldsymbol{\theta}, \boldsymbol{p}, \boldsymbol{q}$ | Voltages, phase angles, active and reactive power of all the buses in the system. |
| $p_{Gi}, q_{Gi}, p_{Di}, q_{Di}$ | Active, reactive power of generators and lumped loads. |
| $pd_i, qg_i$ | Auxiliary variables regarding bus $i$. |
| $G_{ij}, B_{ij}, p_{ij}$ | The $(i, j)$ element of the system's conductance and susceptance matrices; the active power through line $ij$. |
| $p_{w1,0}, \Delta_{w1}$ | Forecasted POI active power and possible deviation. |
| $\alpha_i$ | The participation factor for the generator at bus $i$. |
| $\boldsymbol{\xi}, \boldsymbol{\pi}$ | The vectors of uncertain parameters and control variables for this problem (note that $\boldsymbol{\pi}$ is in bold italics). |
| $\boldsymbol{a}_0, \boldsymbol{a}_\pi, \boldsymbol{a}_\xi, \boldsymbol{a}_{\xi\pi}, \gamma$ | The coefficient matrices and vectors used to explain Soyster's method. |

Note that, due to space limitations, the other *ad hoc* notations will be explained where they first appear. Moreover, unless particularly specified, notations ■̄ and ■ denote the upper and lower limits for a variable placeholder ■, [■$_i$] a vector composed of a series of variable ■$_i$, **1** (in bold) an all-one vector, superscript T transpose.

## I. Introduction

WHIS conventional power plants replaced by renewable generation, power grids in many places, such as Germany, are short of reactive power resources [1]-[3]. In this context, it has been confirmed that doubly-fed induction generators (DFIGs) can provide reactive power support to power grids [4]-[6]. For example, a recent pilot project in Lolland shows that the wind farms there "offered the required amount of 40 Mvar reactive power" [6]. Hence, although the issue of economic incentives remains to be solved [6],[7], it is fair to say that leveraging reactive power potential (RPP), which is a deterministic and continuous range of the adjustable reactive power [2], from wind turbines and wind farms, is promising for power systems that lack reactive power resources.

For the system operator to leverage this RPP, there are two paradigms in general: one is to interface with every DFIG in

This work was supported in part by the National Natural Science Foundation of China under Grant 52007105, Young Elite Scientists Sponsorship Program by CSEE under Grant JLB-2020-170, and Qilu Youth Scholar Program from Shandong University.

Yu Zhou and Zhengshuo Li are with the School of Electrical Engineering, Shandong University, Jinan 250061, China. Zhengshuo Li is the corresponding author (email: zsli@sdu.edu.cn).

wind farms directly, and the other is to interact with a whole wind farm. This paper suggests and focuses on the second paradigm because of the following considerations [8],[9]: (1) it would simplify system operators' supervision and control process and thus probably be more practicable and scalable[1], and (2) the issue of data privacy - wind farms may be reluctant and even resist to let the system operator know its inner model and data.

Although leveraging the RPP of a whole wind farm is more attractive, it requires a wind farm control center to accurately evaluate the RPP, which is technically challenging because of (1) nonlinear power flow equations regarding a wind farm, (2) uncertain parameters such as DFIGs' active power, and (3) discrete controls in a wind farm, such as switchable capacitor banks and online load tap changers (OLTCs) [10],[11]. In the literature, how to accurately and efficiently evaluate wind farm RPP has yet to be addressed. Ref. [7] presents a wind farm RPP evaluation method considering DFIGs' uncertain active power. However, the researchers neglected the discrete controls and the uncertainty in the voltage setpoint of the point of interconnection (POI), which is subsequently set by the system operator through an OPF program after receiving the RPP[2]. Ref. [10] considers the uncertainty and presents a nonlinear RPP evaluation method for a wind farm based on mixed-integer second-order cone programming (MISOCP). Since solving MISOCP is typically computationally expensive for large-scale problems, this method, though accurate, might be inefficient for online applications. A similar method was also applied to evaluate the RPP of a distribution system [12]. To lighten the computational burdens, the researchers then tested a linear RPP evaluation method in [13] by using the linear DistFlow equations [14],[15]; however, since the network losses regarding the distribution system are ignored in this set of linear DistFlow equations, this method might yield an inaccurate RPP evaluation. Hence, one needs to balance the accuracy and computational cost properly when evaluating RPP.

In addition to the RPP evaluation problem that plagues wind farm control centers, the upstream system operator, after receiving wind farm RPP, also faces challenges to her/his RPP-involved reactive power optimization (RPO) process. One major challenge to be considered in this paper is that this RPO can be afflicted with uncertainty like uncertain lumped loads and renewable generation [16],[17]. In the literature, these uncertainties are often modeled as uncertain deviations from a deterministic forecast value (defined as *base-case value*) that will occur most likely, and these deviations are typically assumed to follow a certain type of probability distributions or stay within an interval, e.g., [16]-[19]. To improve this RPP-involved RPO's practicability, we will consider these uncertainties and study the model and solution method for this uncertain RPO problem.

To summarize, this paper suggests that a system operator in need of reactive power resources leverage the wind farm RPP in her/his RPO process. To end this, we present two methods: one for a wind farm control center to evaluate wind farm RPP and one for system operators to optimize the usage of RPP and other dispatchable resources in an uncertain RPO problem. Compared with previous relevant studies, the technical contribution of this paper is twofold:

1) To balance computational accuracy and cost of evaluating RPP, we use a simplified DistFlow model that linearly approximates network losses. On the one hand, this method has the same computational expense order as the method in [13] but is more accurate. On the other hand, although not as accurate as the MISOCP-based method in [10],[12], this method can be solved much more efficiently. This advantage will be substantiated in case studies. Moreover, we also introduce a preference weight in this RPP evaluation method to reflect a system operator's preference for inductive or capacitive RPP in the subsequent RPO process.

2) Unlike our previous study [10],[12] that neglect the uncertainty in the RPO process, we consider these uncertainties and establish an "*AC-base and linear-superposition*" procedure to ensure system-wide power flow feasibility under both the *base case* (the case where the uncertain parameters are set to the base-case values) and any realization of uncertain deviations from the base-case values. The effect of automatic generation control (AGC) and local voltage control against the deviations are also captured in our uncertain RPO model.

In addition to the above contribution, this paper also demonstrates through case studies that wind farms, despite uncertainty in the DFIGs' active power, are still *competent* reactive power resources for system operators, and the safety and optimality of system operations are thus notably improved.

The remainder of the paper is arranged as follows. Section II delineates the proposed linear RPP evaluation method on top of a qualitative analysis of three interaction approaches. Section III shows the model and the solution method regarding the uncertain RPO problem. Section IV presents the case study results and related analysis. Conclusions are given in Section V.

## II. WIND FARM REACTIVE POWER POTENTIAL

### A. Three Interaction Approaches

Below we briefly present three approaches in which a wind farm may interact with a power system on RPP.

*Approach 1*: *Do-not-provide-RPP* ($RPP_0$). In this approach, a wind farm control center fixes the POI reactive power to a constant value (e.g., zero), so there is no room for the upstream system operator to adjust the POI reactive power in the RPO process. This $RPP_0$ represents a conventional and *passive* way in which wind farms interact with power systems [11].

*Approach 2*: *RPP-by-deterministic-methods* ($RPP_D$). In this approach, a wind farm control center evaluates the RPP but neglects the impact of uncertainty, e.g., the DFIG's uncertain active power. The deterministic method of evaluating a distribution system RPP in [20] exemplifies this approach. Detailed mathematical formulations given in [20],[12] have shown that evaluating $RPP_D$ is computationally straightforward. However, if a realization of the uncertainty deviates significantly from the expectation adopted for this evaluation, the resultant $RPP_D$ can be unreliable in the sense that the system operator's required POI reactive power, though located in $RRP_D$, is not actually providable due to the wind farm operational constraints; otherwise, the operational security of this wind farm will be compromised. This is a major drawback of this approach.

---

[1] To see this, imagine two scenarios: in the first the system operator supervises wind farms' reactive power at the points of interconnection (POIs), which is a standard optimal power flow (OPF) problem; in the second the operator has to supervise tens or even hundreds of DFIGs in those wind farms while considering operational constraints regarding every farm. For the cases that multiple wind farms are considered, the latter is extremely complicated.

[2] The reason for modeling the POI voltage setpoint as an uncertain parameter is similar to that for modelling the boundary voltage setpoint in the problem of evaluating the RPP of a distribution system, as explained in [12].

*Approach 3*: *RPP-by-robust-optimization-methods* ($RPP_R$). When considering uncertainty, it is straightforward to see that an RPP range would likely vary with different realizations of uncertainty. For example, the wind turbine's uncertain active power may affect both the adjustable range of this turbine's reactive power and the power flow inside the wind farm, consequently affecting the whole wind farm RPP. Nevertheless, one can expect that these RPP ranges would still have a non-empty *intersection*, denoted by $RPP_R$, which is *constant* against any realization of uncertainty. This expectation can be verified and this $RPP_R$ constructed by solving the following problem:

To find $[\underline{q}_{w1}, \overline{q}_{w1}]$ having the largest span, such that for any POI reactive power setpoint $q_{w1} \in [\underline{q}_{w1}, \overline{q}_{w1}]$, the operational constraints of the wind farm are always feasible for any realization of uncertain parameters.

If this problem has a solution $[\underline{q}_{w1}, \overline{q}_{w1}]$, we can assert that it is possible to construct this $RPP_R$ that is immune to uncertainty.

### B. Construction of $RPP_R$

As shown in the previous studies [10],[12], constructing the $RPP_R$ can be formulated as a two-stage robust optimization (RO) problem through a series of mathematical transformations. However, the model in [12] entails mixed-integer variables as well as an SOC-relaxed DistFlow model, which imposes heavy computational burdens for large-scale problems. Meanwhile, the linear model in [13], though readily solvable, neglects network losses, so it may not always meet our expectation of accuracy.

In this paper, we use a simplified DistFlow model that linearly approximates the losses using the wind farm's real-time measurements $p_{wij,0}, q_{wij,0}$ which are always available in practice. Another difference from the previous study is that in addition to maximizing the span of the $RPP_R$, we introduce a preference weight $\beta$ to the objective function in (1a) to reflect the system operator's preference for inductive or capacitive RPP (larger $\beta$ corresponds to more inductive RPP, and vice versa):

$$\text{minimize } \left(\underline{q}_{w1} - \underline{Q}_w\right)^2 + \beta\left(\overline{q}_{w1} - \overline{Q}_w\right)^2. \quad (1a)$$

Given $p_{wij,0}, q_{wij,0}$, the classical DistFlow equations will be first simplified as follows:

$$p_{wj} = \sum_{k \in \mathcal{T}(j)} p_{wjk} - \sum_{i \in \mathcal{H}(j)} \left(p_{wij} - r_{wij} l_{wij}\right), \quad (1b)$$

$$q_{wj} = \sum_{k \in \mathcal{T}(j)} q_{wjk} - \sum_{i \in \mathcal{H}(j)} \left(q_{wij} - x_{wij} l_{wij}\right), \quad (1c)$$

$$v_{wj} = v_{wi} - 2\left(r_{wij} p_{wij} + x_{wij} q_{wij}\right) + \left(r_{wij}^2 + x_{wij}^2\right) l_{wij}, \quad (1d)$$

$$p_{wij,0} p_{wij} + q_{wij,0} q_{wij} = v_{wi} l_{wij}. \quad (1e)$$

The justification for replacing the original $p_{wij}, q_{wij}$ by $p_{wij,0}, q_{wij,0}$ in (1e) is illustrated in Fig.1 The shadow sector shows that unless the future operating point significantly deviates from the current one, the accuracy of (1e) would be acceptable. Furthermore, use the real-time measurement or the nominal value (e.g., 1.0 p.u.) of $v_{wj}$ to eliminate the remaining nonlinearity in (1e)[3]. Thus far, we have set up a simplified DistFlow model.

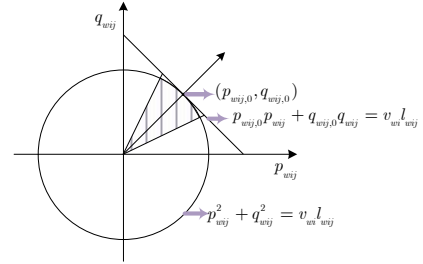

Fig.1. Illustration of the accuracy of (1e) on $p_{wij} - q_{wij}$ plane.

As for a line equipped with an OLTC with $t_{wij}$, (1d) is reformulated using McCormick's convex envelop as follows:

$$v_{wj} = t_{wij}^2 v_{wi} - 2\left(r_{wij} p_{wij} + x_{wij} q_{wij}\right) + \left(r_{wij}^2 + x_{wij}^2\right) l_{wij}, \quad (1f)$$

$$\begin{cases} t_{wij}^2 v_{wi} = \sum_\kappa TR_{wij,\kappa}^2 z_{wTij,\kappa}, \\ v_{wi} - (1 - b_{wTij,\kappa}) \overline{v}_{wi} \leq z_{wTij,\kappa} \leq v_{wi} - (1 - b_{wTij,\kappa}) \underline{v}_{wi}, \\ b_{wTij,\kappa} \underline{v}_{wi} \leq z_{wTij,\kappa} \leq b_{wTij,\kappa} \overline{v}_{wi}, \sum_\kappa b_{wTij,\kappa} = 1, b_{wTij,\kappa} \in \{0,1\}. \end{cases} \quad (1g)$$

The current limit for every line reads $\underline{l}_{wij} \leq l_{wij} \leq \overline{l}_{wij}$.

As for the nodes connected with DFIGs, capacitors and SVGs, the nodal power injections are formulated as:

$$p_{wj} = p_{wGj}, \quad q_{wj} = q_{wGj} + q_{wSj} + q_{wCj}. \quad (1h)$$

The possibly existing static var generators (SVGs) and switchable capacitor banks in the wind farm are modeled as

$$\underline{S}_{wSj} \leq q_{wSj} \leq \overline{S}_{wSj}, \quad (1i)$$

$$\begin{cases} q_{wCj} = \sum_\kappa CP_{wj,\kappa} z_{wCj,\kappa}, \\ v_{wj} - (1 - b_{wCj,\kappa}) \overline{v}_{wj} \leq z_{wCj,\kappa} \leq v_{wj} - (1 - b_{wCj,\kappa}) \underline{v}_{wj}, \\ b_{wCj,\kappa} \underline{v}_{wj} \leq z_{wCj,\kappa} \leq b_{wCj,\kappa} \overline{v}_{wj}, \sum_\kappa b_{wCj,\kappa} \leq 1, b_{wCj,\kappa} \in \{0,1\}. \end{cases} \quad (1j)$$

The active and reactive power of a single DFIG is modeled as follows:

$$p_{wGj}^2 + q_{wGj}^2 \leq S_{wGj}^2, \quad (1k)$$

$$p_{wGj,0} - \Delta_{wGj} \leq p_{wGj} \leq p_{wGj,0} + \Delta_{wGj}. \quad (1l)$$

where (1l) means that the uncertain $p_{wGj}$ may continuously vary between $[p_{wGj,0} - \Delta_{wGj}, p_{wGj,0} + \Delta_{wGj}]$. The

---

[3] Notice that if we use the nominal values of $p_{wij}, q_{wij}, v_{wj}$ rather than the real-time measurements, then this set of simplified DistFlow equations regresses to the equations used in [21].

constraint (1k) can be further linearized as [12], where $K$ can be 4 or 8:

$$-S_{wGj} \leq \cos(\iota\tfrac{\pi}{K})p_{wGj}+\sin(\iota\tfrac{\pi}{K})q_{wGj} \leq S_{wGj}, \ \iota=1,\cdots,K. \quad (1m)$$

Supposing that the nodal voltage square should be constrained in a reasonable range when the wind farm provides RPP, we have

$$v_{w1} = (V_{POI}^{set})^2, \text{ and } \underline{v}_{wj} \leq v_{wj} \leq \overline{v}_{wj}, \ \forall j \neq 1. \quad (1n)$$

As explained in Footnote 2, unlike [22], $V_{POI}^{set}$ in (1n) is taken as an uncertain parameter varying between $[\underline{V}_{POI}^{set}, \overline{V}_{POI}^{set}]$.

In addition to the uncertain $p_{wGj}$ and $V_{POI}^{set}$, the system operator's POI reactive power setpoint, i.e., $q_{w1}$, is also uncertain to the wind farm control center in this RPP evaluation stage, but it should be located within the RPP$_\text{R}$ $[\underline{q}_{w1}, \overline{q}_{w1}]$. To model this, we introduce an uncertain parameter $u_q \in [0,1]$ and formulate $q_{w1}$ as $q_{w1} = \underline{q}_{w1} + u_q(\overline{q}_{w1} - \underline{q}_{w1})$.

Thus far, we have concretely modeled all the variables, constraints, and uncertainty in the RPP evaluation problem. As the variables related to the RPP bounds, the tap positions and on/off status of the switchable capacitors, denoted by $x$, should be determined before checking the feasibility of the wind farm operational constraints under any realization of the uncertainty parameters $u = [[p_{wGj}]; V_{POI}^{set}; u_q]$ [10],[12],[23], the above model is mathematically a two-stage RO problem whose canonical formulation is shown in (2),

$$\min_{x} f(x)$$
$$s.t. \begin{cases} Jx \leq s, & \text{for } \forall u \in \mathcal{U} = \{u | Ru \leq r\}, \\ \exists y: Au + By \leq g, Cu + Dy = h, Ex + Fy \leq m \end{cases} \quad (2)$$

where $y$ denotes the left variables in (1) excluding $x$ and $u$, $f$ the formula in (1a), $\mathcal{U}$ the polyhedral uncertainty set constrained by the aforementioned limits on $p_{wGj}, V_{POI}^{set}, u_q$. Although (2) looks similar to the evaluation model in [13], it should be noted that they are physically different because network losses are considered here.

Nevertheless, due to the similarity, one can still adopt the algorithm in [13] to solve (2). Its flowchart is shown in Fig. 2. Compared with the original column-and-constraint generation (C&CG) algorithm in [24], our algorithm is a reduced version: no need to record and compare upper and lower bounds to terminate. This is both because the second-stage problem in (2) is to check whether there exists a $y$ for any $u$ for a given $x$, which corresponds to $\mathcal{Q}_{k+1} = 0$ in Fig. 2 [13], and because $f(x)$, the objective function in both (2) and the master problem in Fig. 2, is increasingly constrained with more constraints added during the iteration. Moreover, it is not hard to see that the convergence of this reduced C&CG algorithm depends on a finite number of the extreme points of the uncertainty set $\mathcal{U}$, like the algorithm in [24]. Hence, it will also converge within finite iterations as long as $\mathcal{U}$ is a polyhedron or a finite discrete set.

### III. RPP-INVOLVED REACTIVE POWER OPTIMIZATION

#### A. Statement of Problem

As stated previously, one advantage of leveraging the whole wind farm RPP is to simplify system operators' RPO process. For a system operator knowing every wind farm RPP $[\underline{q}_{w1}, \overline{q}_{w1}]$, s/he only needs to solve an OPF problem to determine the usage of the RPP and other dispatchable resources, e.g., for minimal network losses, and then sends the wind farm control centers the commands of how much reactive power is required from the wind farm, i.e., the POI reactive power setpoint $q_{w1}$, as well as the POI voltage setpoint $V_{POI}^{set}$. Then, the wind farm control center disaggregates this $q_{w1}$ among the wind turbines (and SVCs if any). This framework is shown in Fig. 3. Moreover, the wind farm should inform the system of its POI active power forecast value $p_{w1,0}$ and the radius of its possible deviation $\Delta_{w1}$. The method of forecasting $p_{w1,0}$ and estimation of $\Delta_{w1}$ is beyond the scope of this paper. Interested readers can refer to [25],[26].

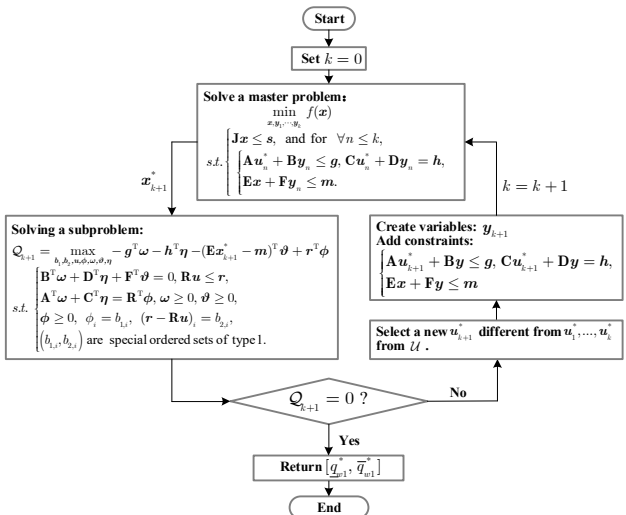

Fig.2. Flowchart of the algorithm to solve (2).

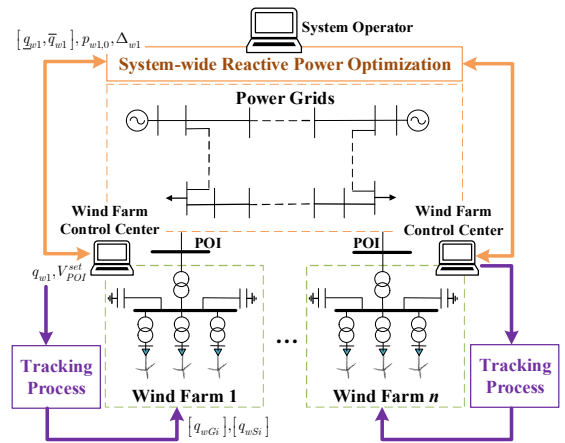

Fig.3. The schematic diagram for the wind farm RPP-involved RPO process.

In this RPO stage, the system operator may face uncertainty in lumped loads as well as the aggregated active power of wind farms (e.g., represented by $p_{w1,0}, \Delta_{w1}$) and solar stations. Hence, the security regarding the base case and any possible scenario should be ensured in this RPO, which is technically challenging. To overcome this challenge, we establish an *"AC-base and linear-superposition"* procedure, which imitates a commonly adopted real-time dispatch paradigm that an operator determines base-case setpoints to ensure system-wide AC-power-flow feasibility while leveraging AGC and local voltage control against uncertain deviations from the base case. It should be noted that this procedure relies on a common and usually tenable assumption that the deviations should stay within a reasonable region such that linearized power flow equations would be a good approximation of the relationships between power injection and bus voltages.

### B. Mathematical Model

*1) Feasibly constraints for the base case*

For every bus in the base case, we have the following AC power flow constraints:

$$p_i = p_{Gi} - pd_i = V_i \sum_j V_j \left( G_{ij} \cos\theta_{ij} + B_{ij} V_j \sin\theta_{ij} \right), \quad (3a)$$

$$q_i = qg_i - q_{Di,0} = V_i \sum_j V_j \left( G_{ij} \sin\theta_{ij} - B_{ij} V_j \cos\theta_{ij} \right), \quad (3b)$$

where $pd_i = p_{Di,0} - \sum_{w \in i} p_{w1,0}$, $qg_i = q_{Gi} + \sum_{w \in i} q_{w1}$, $\theta_{ij} = \theta_i - \theta_j$, and $w \in i$ denote all the wind farms connected to bus $i$, the subscript 0 the forecasted base-case value. The POI setpoint for the wind farm at bus $i$ is set as $V_{POI}^{set} = V_i$.

The line flow, bus voltages, and the controls should stay within their limits in this base case:

$$\underline{p}_{ij} \leq p_{ij} = G_{ij}(V_i V_j \cos\theta_{ij} - V_i^2) + B_{ij} V_i V_j \sin\theta_{ij} \leq \overline{p}_{ij}, \quad (3c)$$

$$\underline{V}_i \leq V_i \leq \overline{V}_i, \ \underline{p}_{Gi} \leq p_{Gi} \leq \overline{p}_{Gi}, \ \underline{q}_{Gi} \leq q_{Gi} \leq \overline{q}_{Gi}, q_{w1} \in [\underline{q}_{w1}, \overline{q}_{w1}]. \quad (3d)$$

*2) Feasibly constraints against deviations*

When a realization of uncertainty deviates from the base case to some extent, AGC and local voltage control will be activated. Accurate emulation of these dynamic control behaviors is complicated and beyond the scope of this paper. Instead, we capture the steady-state behaviors of these controls, e.g., the linear control policy of the AGC on top of a linear power flow model.

*a) Linear power flow model*

Classify the system buses into four categories: (i) a reference bus whose voltage remains constant and phase angle = 0; (ii) a group of AGC buses connected to AGC generators that are assumed to be equipped with local voltage controls as well; (iii) voltage control buses that do not connect with AGC generators but have voltage control facilities, e.g., wind farms providing RPP with uncertain generation (see Fig. 3); (iv) the remaining buses with no AGC nor voltage controls. Let subscripts $\mathcal{R}$, $\mathcal{A}$, $\mathcal{V}$, and $\mathcal{D}$ denote these four bus categories respectively, and the prefix $\Delta$ the deviation or increment of the variables. Then, we have $\Delta\theta_{\mathcal{R}} = 0$, and it follows from the linear power flow model in [27] that

$$\begin{bmatrix} \Delta\boldsymbol{\theta}_{\mathcal{A}\cup\mathcal{V}\cup\mathcal{D}} \\ \Delta\boldsymbol{V} \end{bmatrix} = \mathbf{L} \begin{bmatrix} [\Delta\boldsymbol{p}_{\mathcal{R}\cup\mathcal{A}}; \Delta\boldsymbol{p}_{\mathcal{V}\cup\mathcal{D}}] \\ [\Delta\boldsymbol{q}_{\mathcal{R}\cup\mathcal{A}\cup\mathcal{V}}; \Delta\boldsymbol{q}_{\mathcal{D}}] \end{bmatrix}, \quad (4a)$$

where $\mathbf{L}$ is a constant matrix.

*b) Active power deviation and AGC model*

Without loss of generality, let $\Delta\boldsymbol{d} \in [\Delta\underline{\boldsymbol{d}}, \Delta\overline{\boldsymbol{d}}]$ be the active power deviations caused by uncertain loads and renewable generation[4] at all buses, so we have $\Delta\boldsymbol{p}_{\mathcal{V}\cup\mathcal{D}} = \Delta\boldsymbol{d}_{\mathcal{V}\cup\mathcal{D}}$. The total deviation amount $\mathbf{1}^T\Delta\boldsymbol{d}$ should be offset by the AGC system that is assumed to take a linear control policy with non-negative participation factor vector $\boldsymbol{\alpha}$ regarding the AGC generator as well as the "virtual" generator at the reference bus to be optimized. Specifically, $\Delta p_{Gi} = (\mathbf{1}^T\Delta\boldsymbol{d})\alpha_i$, so $\Delta\boldsymbol{p}_{\mathcal{R}\cup\mathcal{A}} = (\mathbf{1}^T\Delta\boldsymbol{d})\boldsymbol{\alpha} - \Delta\boldsymbol{d}_{\mathcal{R}\cup\mathcal{A}}$ subject to the following technical constraints:

$$\mathbf{1}^T\boldsymbol{\alpha} = 1, \ \boldsymbol{\alpha} \geq 0, \ \underline{p}_{Gi} \leq p_{Gi} + \Delta p_{Gi} \leq \overline{p}_{Gi}. \quad (4b)$$

*c) Reactive power deviations and local voltage control model*

First, assume the reactive power deviations for the category $\mathcal{D}$ is $\Delta\boldsymbol{q}_{\mathcal{D}} = \boldsymbol{\sigma}\Delta\boldsymbol{d}_{\mathcal{D}}$, where $\boldsymbol{\sigma}$ is the diagonal matrix of the power factors regarding the related buses. Next, following the idea in [28] one can derive linear incremental relationships between local reactive power and voltage magnitudes as follows:

$$\Delta\boldsymbol{q}_{\mathcal{R}\cup\mathcal{A}\cup\mathcal{V}} = \mathbf{M}[\Delta\boldsymbol{p}_{\mathcal{A}\cup\mathcal{V}\cup\mathcal{D}}; \Delta\boldsymbol{q}_{\mathcal{D}}] + \mathbf{N}\Delta\boldsymbol{V}_{\mathcal{R}\cup\mathcal{A}\cup\mathcal{V}}, \quad (4c)$$

where $\mathbf{M}$ and $\mathbf{N}$ are constant matrices. To further simplify (4c), which would otherwise make this RPO a difficult two-stage nonlinear optimization problem, we assume $\Delta\boldsymbol{V}_{\mathcal{R}\cup\mathcal{A}\cup\mathcal{V}} \approx 0$ based on the fact that local voltage control typically makes local voltages deviate insignificantly from the base case [29]. Hence, (4c) is further simplified as

$$\Delta\boldsymbol{q}_{\mathcal{R}\cup\mathcal{A}\cup\mathcal{V}} = \mathbf{M}[\Delta\boldsymbol{p}_{\mathcal{A}\cup\mathcal{V}\cup\mathcal{D}}; \boldsymbol{\sigma}\Delta\boldsymbol{d}_{\mathcal{D}}]. \quad (4d)$$

The reactive power regulated as (4d), i.e., a function of uncertainty $\Delta\boldsymbol{d}$, $\boldsymbol{\sigma}\Delta\boldsymbol{d}_{\mathcal{D}}$ and control variables $\boldsymbol{\alpha}$, should ensure that the system-wide bus voltage magnitudes stay within their acceptable regions $[\underline{V}_i, \overline{V}_i]$ with voltage control facilities and wind farms constrained by their technical limits as follows,

$$\underline{V}_i \leq V_i + \Delta V_i \leq \overline{V}_i, \ \underline{q}_{Gi} \leq q_{Gi} + \Delta q_{Gi} \leq \overline{q}_{Gi}, q_{w1} + \Delta q_{w1} \in [\underline{q}_{w1}, \overline{q}_{w1}]. \quad (4e)$$

*d) Other operational constraints*

We can formulate other operational constraints against deviation scenarios, e.g., the line's power flow limits below:

$$\underline{p}_{ij} \leq p_{ij} + \Delta p_{ij} \leq \overline{p}_{ij} \quad (4f)$$

where $\Delta p_{ij} = -G_{ij}(\Delta V_i - \Delta V_j) + B_{ij}(\Delta\theta_i - \Delta\theta_j)$ [27]. This constraint means that the power flow through every line should be limited by its transmission capacity[5].

*3) Integral Model*

Thus far, we have modeled the feasibility constraints for both the base case in (3) and any possible deviation scenario in (4). If we choose minimizing the system-wide network losses in the base case[6], as shown in (5),

$$\min_{\boldsymbol{\pi}} \sum G_{ij}[V_i^2 + V_j^2 - 2V_i V_j \cos(\theta_i - \theta_j)], \quad (5)$$

as the objective, we finally formulate the following uncertain RPO problem:

---

[4] For instance, as for a wind farm, its $[\Delta\underline{d}, \Delta\overline{d}] = [-\Delta_{w1}, \Delta_{w1}]$.
[5] Since instantaneous three-phase power equals the active power, lines' thermal effect is usually formulated through active power constraint [30].
[6] The choice of objective functions is not unique: one can add the regulation cost of the AGC, which is also a function of control $\boldsymbol{\pi}$, into (5). In general, the choices will not affect workability of the following solution method.

*Objective function:* (5)  *Constraints:* (3) and (4)
*Uncertainty* $\boldsymbol{\xi} = [\Delta \boldsymbol{d}; \Delta \boldsymbol{q}_{\mathcal{D}}]$  *Control* $\boldsymbol{\pi} = [[p_{Gi}]; [q_{Gi}]; [q_{w1}]; \boldsymbol{\alpha}]$

*C. Solution Method*

Substituting (4a) and (4d) into (4b), (4e), and (4f), one can derive a set of inequalities linear in $\boldsymbol{\xi}$, $\boldsymbol{\pi}$, and/or $(\boldsymbol{a}_{\xi\pi}\boldsymbol{\pi})^T\boldsymbol{\xi}$. Since the uncertainty set of $\boldsymbol{\xi}$ is box-shaped for this RPO problem, one can adopt Soyster's method in [31] to convert these constraints affected by $\boldsymbol{\xi}$ into their equivalent deterministic counterpart. To show this, a canonical formula representative of every inequality in (4) is first given below,

$$\max_{\boldsymbol{\xi}\in[\underline{\boldsymbol{\xi}},\overline{\boldsymbol{\xi}}]}\{a_0 + \boldsymbol{a}_\pi^T\boldsymbol{\pi} + (\boldsymbol{a}_\xi + \boldsymbol{a}_{\xi\pi}\boldsymbol{\pi})^T\boldsymbol{\xi}\} \leq \gamma . \quad (6a)$$

Soyster's method introduces an auxiliary variable $\lambda$ to equivalently transform (6a) to a set of linear inequalities as follows:

$$\begin{cases} (\boldsymbol{a}_{\xi\pi}^T\underline{\boldsymbol{\xi}} + \boldsymbol{a}_\pi)^T\boldsymbol{\pi} + \lambda \leq \gamma - a_0 - \boldsymbol{a}_\xi^T\underline{\boldsymbol{\xi}} , \\ \lambda - (\overline{\boldsymbol{\xi}} - \underline{\boldsymbol{\xi}})^T\boldsymbol{a}_{\xi\pi}\boldsymbol{\pi} \geq \boldsymbol{a}_\xi^T(\overline{\boldsymbol{\xi}} - \underline{\boldsymbol{\xi}}), \lambda \geq 0. \end{cases} \quad (6b)$$

Using the transformation in (6), one can transform the uncertain RPO problem into a deterministic nonlinear optimization problem, which is readily solvable by off-the-shelf solvers through an interior-point method [32],[33]. We should point out that in this RPO problem, integer variables are not explicitly considered. Because of the complexity of the this RPO formulation, if the system-wide discrete controls need to be involved in the RPO process, we suggest adopting a heuristic approach in which the integers are first taken as continuous variables for a temporary RPO solution and then refined heuristically, e.g., with a classical round-off approach. Interested readers can refer to [34].

*D. Tracking Reactive Power Command*

Consistent with the assumptions for the two-stage-RO-based RPP evaluation model, the uncertainty regarding the wind farm has been realized when the control center tracks the system operator's reactive power command, i.e., the POI reactive power setpoint $q_{w1}$. Hence, the wind farm control center only needs to solve a regular OPF problem to disaggregate this $q_{w1}$ among the wind turbines and SVCs if any, namely, to specify the reactive power provided by every wind turbines and SVC, i.e., $q_{wGj}$ and $q_{wSj}$ (see Fig. 3). The constraint set of this tracking problem includes (1b)–(1n) (here (1e) can recover its nonlinearity) with the discrete variables fixed to the values solved from (2) for the reason explained in [13],[23]. The objective function can be minimizing the deviations of the tracking as [13], or together with other optional terms such as minimizing the network losses inside the wind farm [22],[35]. More details about this tracking process are given in [13].

## IV. CASE STUDIES

*A. Simulation Systems*

The first simulation was performed on a small-scale system for conceptual validation. We adopted an IEEE 9-bus system whose original load at bus 5 was replaced by a wind farm [36]. This configuration is shown in Fig. 4. The four bus categories for this system are $\mathcal{R}$:{Bus 1}, $\mathcal{A}$: {Buses 2, 3}, $\mathcal{V}$: {Bus 5}, and $\mathcal{D}$: {the remaining buses}. The voltage magnitude limits at Bus 5, namely the wind farm POI, is set to [0.99, 1.01] p.u. The voltage magnitude limits regarding the other buses are set to [0.97, 1.03] p.u. The adjustable range regarding all the five-bus-system generator active power is set to [40, 100] MW. The base-case values of the uncertain lumped loads are set to the values specified in the case-data file in [37], and the uncertain deviations are up to $\pm 10\%$ of the base-case values. As for the wind farm, there are six DFIGs. The OLTC at Line W1-W2 is adjustable in the range [0.98, 1.02] with a step 0.01. The transformer at Line W3-W4 is not adjustable with the ratio set to 1. The capacitor shunts at Nodes W3 and W4 have two switchable 2.5-Mvar banks individually. As for each wind turbine, we assume that $p_{wGj,0}$ is 1.0 MW, $\Delta_{wGj}$ is 0.1 MW. The wind farm's POI active power forecast and the radius of the uncertain deviation are taken as the sum of these $p_{wGj,0}, \Delta_{wGj}$[7]. The other data for the system parameters and wind farms can be referred to [36],[37].

The second simulation was performed on a larger system. We adopted an IEEE 39-bus system where its original load at Buses 10 and 17 are replaced by ten similar wind farms, respectively. Since the ten wind farms at the same bus share the same POI, the RPP, the active power forecast, and radius of the deviations at this bus are the summation of the counterparts of the wind farms. The settings of the POI voltage magnitude limits and uncertain lumped loads are the same as the first simulation. Moreover, $\mathcal{R}$ includes the reference bus, $\mathcal{A}$ the PV buses, $\mathcal{V}$ Buses 10 and 17, and $\mathcal{D}$ the remaining buses. The other data are also available in [36] and [37]. In the following tests, $\beta$ in (1a) is set to 1.0 by default, representing system operators' neutral preference for capacitive or inductive RPP.

MATLAB and GUROBI are used as the simulation environment and the solver, respectively. The codes are implemented on a PC with an Intel i5 CPU@3.0 GHz and 8 GB of RAM.

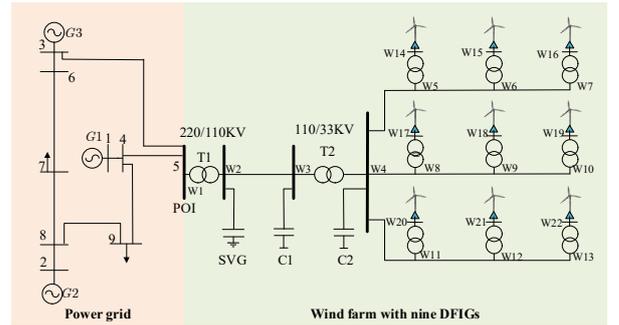

Fig. 4. Graphic of the modified IEEE 9-bus system.

*B. Tests on RPP Evaluation Methods*

*1) Accuracy and computational efficiency*

On the 9-bus system with one wind farm, we compare the computational accuracy and efficiency regarding four RPP evaluation methods: $RPP_D$ in [20] and [12], a nonlinear $RPP_R$ ($RPP_{R-1}$) in [10],[12], a linear $RPP_R$ ($RPP_{R-2}$) neglecting network losses in [13], and the method proposed in this paper (RPP$_{R-3}$). The results are shown in Table I. First, it shows that RPP$_D$ is larger than the three RPP$_R$'s, both because the former ignores the uncertainty, which may risk the following system-wide RPO as discussed in Section II.A, and because RPP$_D$ employs inconsistent optimal capacitor banks (i.e., ([5,5], [0,0]) and tap positions (i.e., 0.98, 1.02) for $\underline{q}_{w1}, \overline{q}_{w1}$, which may not be practically attainable [23]. Therefore, though evaluating RPP$_D$ is fastest in this test, it may yield risky RPP. Second, Table I shows that RPP$_{R-1}$ is accurate but requires the longest computing time, several times of that of RPP$_{R-2}$ and RPP$_{R-3}$. This deficiency would be more severe for larger systems. Third, in comparison with RPP$_{R-2}$, RPP$_{R-3}$ is closer to RPP$_{R-1}$ because network losses are approximately considered in this method, but their computing times are of the same order.

---

[7] We adopted this rough estimation to simplify the simulation. Accurate estimation methods as noted in Section III.A are beyond the scope of this paper.

TABLE I
COMPARISON OF FOUR RPP EVALUATION METHODS

| Approaches | RPP$_D$ (terms related to $\underline{q}_{w1}, \overline{q}_{w1}$) | RPP$_{R-1}$ | RPP$_{R-2}$ | RPP$_{R-3}$ |
|---|---|---|---|---|
| RPPs (Mvar) | [-18.32,15.31] | [-7.17,5.84] | [-7.57,5.37] | [-7.32,5.46] |
| Capacitors (Mvar) | [5,5], [0,0] | [0,5] | [0,5] | [0,5] |
| Tap Ratios (p.u.) | 0.98, 1.02 | 1 | 1 | 1 |
| Computing Time (s) | 0.031, 0.036 | 42.05 | 4.52 | 5.06 |

*2) Impact of the uncertainty of DFIG's active power*

Another interesting question is how much the DFIG's uncertain active power would affect RPP$_R$, or would this RPP$_R$ diminish with increasing uncertainty? To answer it, Table II presents RPP$_{R-3}$ with increasing uncertainty measured in terms of the percentage of the radius of uncertain deviations $\Delta_{wGj}$ to the base-case forecast $p_{wGj,0}$. It shows that although this RPP diminishes gradually, a wind farm still provides a considerable RPP range: a span of nearly 11 Mvar even when $\Delta_{wGj}/p_{wGj,0}$ reaches 80%, a notable uncertainty level.

TABLE II
IMPACT OF UNCERTAINTY IN DFIG' ACTIVE POWER

| $\Delta_{G,j}/p_{G0,j}$ | 10% | 20% | 60% | 80% |
|---|---|---|---|---|
| Span of RPP$_{R-3}$ (Mvar) | 12.8 | 12.6 | 11.6 | 11.0 |

*3) Impact of the preference weight*

Three $\beta$'s, 0.5, 1.0, and 1.5, representing three types of system operator's preferences for RPP: capacitive-favored, neutral, and inductive-favored. The simulation results are given in Table III. These results confirm that regulating $\beta$ is a practical approach of providing capacitive- or inductive-skewed wind farm RPP, as indicated by (1a). In real-world applications, this $\beta$ can be dynamically regulated based on the system operator's real-time preference for reactive power.

TABLE III
THE IMPACT OF PREFERENCE WEIGHT ON RPP

| $\beta$ | RPP$_R$ (Mvar) | Capacitors (Mvar) | Tap Ratios (p.u.) |
|---|---|---|---|
| 0.5 | [-16.26 -2.99] | [5, 5] | 0.98 |
| 1 | [-7.32,5.46] | [0, 5] | 1 |
| 1.5 | [-4.82,7.86] | [2.5,0] | 1.01 |

*C. Tests on Leveraging Wind Farm RPP in RPO Process*
*1) Scalability*

As we claimed previously, it is computationally more scalable for system operators to leverage the whole wind farm RPP instead of interfacing with a group of DFIGs. Table IV substantiates this claim. For this modified 39-bus system, the numbers of the RPP problem's variables and constraints are reduced by about 90% by leveraging wind farm RPP. Wind farm control centers undertake most of these saved computational burdens in their RPP evaluation and tracking process, which can be further implemented distributedly among these wind farms.

TABLE IV
NUMBERS OF VARIABLES AND CONSTRAINTS FOR 39-BUS SYSTEM'S RPO

| Interface Objects | Discrete Variables | Continuously Variables | Constraints |
|---|---|---|---|
| With Wind Farms | 0 | 77 | 403 |
| With DFIGs | 60 | 914 | 4183 |

*2) Improvements in system operations*

To validate the effect of leveraging wind farm RPP, Table V compares the optimal values of the 9-bus system's uncertain RPO problem through two RPP approaches, RPP$_0$ and RPP$_{R-3}$. In RPP$_0$, i.e., *do-not-provide-RPP*, which provides a benchmark, the POI reactive power is set to zero. Three system-wide uncertainty levels, measured by the percentage of maximum possible deviation to its base-case value across the system, were simulated. First, for the uncertainty levels 5% and 10%, Table V shows that the system-wide network losses are reduced by about 0.03 MW when the wind farm provides RPP$_R$. Second, it is interesting to see that if this wind farm does not provide RPP, the resultant RPO problem has no solution when the uncertainty level reaches 14%. This test, though performed on this small-scale system, resembles a real-world predicament that with a decreasing amount of conventional reactive power resources, it becomes increasingly difficult to secure power systems, which is plaguing system operators in many places.

TABLE V
THE 9-BUS SYSTEM-WIDE NETWORK LOSSES WITH TWO RPP APPROACHES

| Uncertainty Levels | 5% | 10% | 14% |
|---|---|---|---|
| Losses with RPP$_0$ | 1.88 MW | 1.90 MW | No Solution |
| Losses with RPP$_{R-3}$ | 1.85 MW | 1.87 MW | 1.91 MW |

Furthermore, Table VI compares the network losses for the 39-bus system with 20 wind farms. It can be seen that with multiple wind farms actively providing RPP, the system-wide network losses are reduced by about 7 MW, i.e., 13%. This significant reduction confirms the positive and notable effect of leveraging wind-farm RPP on improving operating optimality, especially for power systems with high wind penetration.

TABLE VI
THE 39-BUS SYSTEM-WIDE NETWORK LOSSES WITH TWO RPP APPROACHES

| Approaches | RPP$_0$ | RPP$_{R-3}$ |
|---|---|---|
| Network Losses (MW) | 52.69 | 45.52 |

V. CONCLUSION

This paper suggests leveraging wind farm RPP in the power system RPO process to improve system operational safety and optimality. First, an RO-based linear RPP evaluation method is proposed, which, as shown by case studies, achieves better accuracy than the previous linear evaluation method and enjoys cheaper computational cost than the nonlinear evaluation method. Case studies also show that a wind farm with rather significant uncertainty in DFIGs still provides a considerable RPP range. System operators' desire for inductive or capacitive reactive power resources can also be satisfied.

Second, an uncertain RPP-involved RPO problem is introduced, formulated, and transformed into a regular optimization problem that can be readily solved. Case studies confirm that with increasing system-wide uncertainty, leveraging wind farm RPP is not only a powerful approach to decreasing network losses but also necessary to secure system operations. This simulation might be a useful indicator for power systems connecting with considerable wind farms.

Future work may include (1) investigating a more accurate linear power flow model in the linear RPP evaluation, (2) exploring a less conservative uncertain RPO model by relaxing the assumption $\Delta V_{\mathcal{R}\cup\mathcal{A}\cup\mathcal{V}} \approx 0$, and (3) collaborating with industry partners for field implementation of this proposal.

REFERENCES

[1] IEEE PES Task Force, "Contribution to bulk system control and stability by distributed energy resources connected at distribution network," IEEE Power & Energy Society, Tech. Rep. PES-TR22, Jan. 2017.
[2] P. Goergens, F. Potratz, M. Gödde, and A. Schnettler, "Determination of the potential to provide reactive power from distribution grids to the transmission grid using optimal power flow," *Int. Univ. Power Eng. Conf. (UPEC)*, Stoke on Trent, 2015, pp. 1–6.
[3] H. Barth, D. Hidalgo, A. Pohlemann, M. Braun, L. H. Hansen and H. Knudsen, "Technical and economical assessment of reactive power provision from distributed generators: Case study area of East Denmark," *EEE Grenoble Conf.*, Grenoble, 2013, pp. 1-6.
[4] F. M. Hughes, O. Anaya-Lara, N. Jenkins and G. Strbac, "Control of DFIG-based wind generation for power network support," *IEEE Trans. Power Syst.*, vol. 20, no. 4, pp. 1958-1966, Nov. 2005.
[5] K. Wang, Y. Wang, Z. Cheng, L. Liu, L. Jia and Y. Liang, "Research on Reactive Power Control of the Grid-Side Converter of DFIG Based Wind Farm," *IEEE Conf. Energy Internet and Energy Syst. Integr. (EI2)*, Beijing, 2018, pp. 1-4.
[6] Ramboll Group, "Ancillary services from new technologies," Tech. Rep. Dec. 2019.
[7] N. R. Ullah, K. Bhattacharya, and T. Thiringer, "Wind farms as reactive power ancillary service providers—Technical and economic issues," *IEEE Trans. Energy Convers.*, vol. 24, no. 3, pp. 661–672, Sep. 2009.
[8] H. Sun, Q. Guo, J. Qi, V. Ajjarapu, R. Bravo, J. Chow, Z. Li, R. Moghe, E. Nasr-Azadani, U. Tamrakar, G. N. Taranto, R. Tonkoski, G. Valverde, Q. Wu and G. Yang, "Review of challenges and research opportunities for voltage control in smart grids," *IEEE Trans. Power Syst*, vol. 34, no. 4, pp. 2790 - 2801, Jul, 2019.
[9] Q. Guo, H. Sun, B. Wang, B. Zhang, W. Wu, and L. Tang, "Hierarchical automatic voltage control for integration of large-scale wind power: Design and implementation," *Electr. Pow. Syst. Res.*, vol. 120, pp. 234–241, Mar. 2015.
[10] Y. Zhou, and Z. Li, "Robust estimation of reactive power support range of DFIG wind farm," *Autom. Elect. Power Syst.*, in press.
[11] D. S. Stock, A. Venzke, L. Löwer, K. Rohrig and L. Hofmann, "Optimal reactive power management for transmission connected distribution grid with wind farms," *IEEE Innovative Smart Grid Tech. - Asia (ISGT-Asia)*, Melbourne, VIC, 2016, pp. 1076-1082.
[12] Z. Li, J. Wang, H. Sun, F. Qiu, and Q. Guo, "Robust estimation of reactive power for an active distribution system," *IEEE Trans. Power Syst.*, vol.34, no. 5, pp. 3395-3407, Sept. 2019.
[13] Y. Zhou, Z. Li and M. Yang, "A framework of utilizing distribution power systems as reactive power prosumers for transmission power systems," *Int. J. Elect. Power and Energy Syst.*, vol. 121, Oct. 2020.
[14] K. Turitsyn, P. Šulc, S. Backhaus and M. Chertkov, "Distributed control of reactive power flow in a radial distribution circuit with high photovoltaic penetration," *IEEE PES General Meeting*, Providence, RI, 2010, pp. 1-6.
[15] M. N. Faqiry and S. Das, "Distributed bilevel energy allocation mechanism with grid constraints and hidden user information," *IEEE Trans. Smart Grid*, vol. 10, no. 2, pp. 1869-1879, Mar. 2019.
[16] X. Guo, H. Bao and J. Li, "Interval optimal power flow model and its Monte Carlo method for AC/DC hybrid power system with wind power," *IEEE Conf. Ind. Electron. Appl. (ICIEA)*, Xi'an, China, 2019, pp. 873-876.
[17] T. Ding, R. Bo, F. Li, Y. Gu, Q. Guo and H. Sun, "Exact penalty function based constraint relaxation method for optimal power flow considering wind generation uncertainty," *IEEE Trans. Power Syst.*, vol. 30, no. 3, pp. 1546-1547, May 2015.
[18] X. Bai, L. Qu and W. Qiao, "Robust AC optimal power flow for power networks with wind power generation," *IEEE Trans. Power Syst.*, vol. 31, no. 5, pp. 4163-4164, Sept. 2016.
[19] C. Lin, W. Wu, M. Shahidehpour, Y. Guo and B. Wang, "A non-iterative decoupled solution of the coordinated robust opf in transmission and distribution networks with variable generating units," *IEEE Trans. Sustain. Energy*, vol. 11, no. 3, pp. 1579-1588, Jul. 2020.
[20] D. S. Stock, A. Venzke, T. Hennig, and L. Hofmann, "Model predictive control for reactive power management in transmission connected distribution grids," *IEEE PES Asia-Pacific Power and Energy Eng. Conf. (APPEEC)*, Oct. 2016, pp. 419–423.
[21] T. Ding, F. Li, X. Li, H. Sun and R. Bo, "Interval radial power flow using extended DistFlow formulation and Krawczyk iteration method with sparse approximate inverse preconditioner," *IET Gener. Transm. Distrib.*, vol. 9, no. 14, pp. 1998-2006, Oct. 2015.
[22] B. Zhang, P. Hou, W. Hu, M. Soltani, C. Chen, and Z. Chen, "A reactive power dispatch strategy with loss minimization for a DFIG-Based wind farm," *IEEE Trans. Sustain. Energy*, vol. 7, no. 3, pp. 914–923, Jul. 2017.
[23] T. Niu, Q. Guo, H. Sun, Q. Wu, B. Zhang and T. Ding, "Autonomous voltage security regions to prevent cascading trip faults in wind turbine generators," *IEEE Trans. Sustain. Energy*, vol. 7, no. 3, pp. 1306–1316, Jul. 2016.
[24] B. Zeng and L. Zhao, "Solving two-stage robust optimization problems using a column-and-constraint generation method," *Oper. Res. Lett.*, vol. 41, no. 5, pp. 457–461, 2013.
[25] C. Wan, Z. Xu and P. Pinson, "Direct interval forecasting of wind power," *IEEE Trans. Power Syst.*, vol. 28, no. 4, pp. 4877-4878, Nov. 2013.
[26] C. Wan, Z. Xu, P. Pinson, Z. Y. Dong and K. P. Wong, "Probabilistic forecasting of wind power generation using extreme learning machine," *IEEE Trans. Power Syst.*, vol. 29, no. 3, pp. 1033-1044, May 2014.
[27] J. Yang, N. Zhang, C. Kang and Q. Xia, "A state-independent linear power flow model with accurate estimation of voltage magnitude," *IEEE Trans. Power Syst.*, vol. 32, no. 5, pp. 3607-3617, Sept. 2017.
[28] C. Duan, W. Fang, L. Jiang, L. Yao and J. Liu, "Distributionally robust chance-constrained approximate AC-OPF with Wasserstein metric," *IEEE Trans. Power Syst.*, vol. 33, no. 5, pp. 4924-4936, Sept. 2018.
[29] P. Li, Q. Wu, M. Yang, Z. Li and N. Hatziargyriou, "Distributed distributionally robust dispatch for integrated transmission-distribution systems," *IEEE Trans. Power Syst.*, in press, early access DOI: 10.1109/TPWRS.2020.3024673.
[30] A. R. Bergen, V. Vittal. *Power Systems Analysis*. 2nd ed. New Jersey: Prentice Hall, 1999.
[31] A. L. Soyster, "Technical note—convex programming with set-inclusive constraints and applications to inexact linear programming," *Oper. Res.*, vol. 21, no. 5, pp. 1154-1157, Oct. 1973.
[32] J. A. Momoh, M. El-Hawary, and R. Adapa, "A review of selected optimal power flow literature to 1993. Part II: Newton, linear programming and interior point methods," *IEEE Trans. Power Syst.*, vol. 14, no. 1, pp. 105–111, Feb. 1999.
[33] D. Kourounis, A. Fuchs, and O. Schenk, "Toward the next generation of multiperiod optimal power flow solvers," *IEEE Trans. Power Syst.*, vol. 33, no. 4, pp. 4005-4014, Jul. 2018.
[34] F. Capitanescu and L. Wehenkel, "Sensitivity-based approaches for handling discrete variables in optimal power flow computations," *IEEE Trans. Power Syst.*, vol. 25, no. 4, pp. 1780-1789, Nov. 2010.
[35] S. Huang, P. Li, Q. Wu, F. Li, F. Rong, "ADMM-based distributed optimal reactive power control for loss minimization of DFIG-based wind farms", *Int. J. Elect. Power and Energy Syst.*, vol. 118, 2020.
[36] K. S. Pandya, J. K. Pandya, S. K. Joshi and H. K. Mewada. "Reactive power optimization in wind power plants using cuckoo search algorithm", in *Metaheuristics and Optimization in Civil Engineering*, Springer, 2016, pp. 181–197.
[37] R. D. Zimmerman, C. E. Murillo-Sánchez, and R. J. Thomas, "MATPOWER: Steady-state operations, planning, and analysis tools for power systems research and education," *IEEE Trans. Power Syst.*, vol. 26, no. 1, pp. 0-19, Feb. 2011.